\documentclass[11pt]{article}
\usepackage[a4paper, margin=1.2in]{geometry}
\usepackage[english]{babel}
\usepackage[T1]{fontenc}  			
\usepackage[utf8]{inputenc}
\usepackage{microtype}			 	
\usepackage{graphicx}
\usepackage{caption}
\usepackage{subcaption}
\usepackage{setspace} 				
\usepackage{mathtools}				
\usepackage{amssymb}
\usepackage{amsfonts}
\usepackage{amsthm}
\usepackage{bm} 					
\usepackage{booktabs}				
\usepackage{array}					%
\usepackage{dcolumn} 				
\usepackage[shortcuts]{extdash} 	
\usepackage[breaklinks=true]{hyperref}
\usepackage{cleveref}
\usepackage[square, sort, numbers, authoryear]{natbib}
\usepackage[dvipsnames]{xcolor}
\usepackage{pgfplots}
\usepackage{authblk}        
\usepackage{verbatim}       
\usepackage{siunitx}
\usepackage{csquotes}
\usepackage{enumitem}

\theoremstyle{theorem}

\theoremstyle{definition}
\newtheorem{defn}{Definition}
\newtheorem{alg}{Algorithm}

\usepackage{physics}				
\usepackage{xparse}					

\DeclareDocumentCommand \student {  }{Student-$t$}
\DeclareDocumentCommand \vb 		{ m }{ \bm{\mathbf{\lowercase{ #1 }}} }
\DeclareDocumentCommand \mb 		{ m }{ \bm{\mathbf{\uppercase{ #1 }}} }
\DeclareDocumentCommand \integral 	{ O{} O{} }{ \int\limits_{#1}^{#2}\! }
\DeclareDocumentCommand \d 			{ O{x} }{\,\mathrm{d}#1}
\DeclareDocumentCommand \T 			{ }{ ^{\top} }
\DeclareDocumentCommand \I 			{ }{ ^{-1} }
\DeclareDocumentCommand \diag 		{ m }{ \mathrm{diag}\pqty{\, #1 \,} }
\DeclareDocumentCommand \eye 		{ }{ \mb{I} }
\DeclareDocumentCommand \vzero	{  }{ \vb{0} }
\DeclareDocumentCommand \N 			{ s O{\vb{x}} O{\vb{0}} O{\eye} }
{
	\IfBooleanTF{#1}{
		#2\sim\mathrm{N}\pqty{ #3,\, #4 }
	}{
	\mathrm{N}\pqty{\left. #2 \,\middle\vert\, #3,\, #4 \right.}
}
}
\DeclareDocumentCommand \St		{ s O{\vb{x}} O{\vb{0}} O{\eye} O{\nu} }
{
	\IfBooleanTF{#1}{
		#2\sim\mathrm{St}\pqty{ #3,\, #4,\, #5 }
	}{
	\mathrm{St}\pqty{\left. #2 \,\middle\vert\, #3,\, #4,\, #5 \right.}
}
}
\DeclareDocumentCommand \G		{ s O{t} O{k} O{\theta} }
{
	\IfBooleanTF{#1}{
		#2\sim\mathrm{G}\pqty{ #3,\, #4}
	}{
	\mathrm{G}\pqty{\left. #2 \,\middle\vert\, #3,\, #4 \right.}
}
}
\DeclareDocumentCommand \IG		{ s O{t} O{k} O{\theta} }
{
	\IfBooleanTF{#1}{
		#2\sim\mathrm{IG}\pqty{ #3,\, #4}
	}{
	\mathrm{IG}\pqty{\left. #2 \,\middle\vert\, #3,\, #4 \right.}
}
}
\DeclareDocumentCommand \E 		{ O{} l m }{ \mathbb{E}_{#1}\bqty#2{ #3 } }
\DeclareDocumentCommand \V 		{ O{} l m }{ \mathbb{V}_{#1}\bqty#2{ #3 } }
\DeclareDocumentCommand \Cov 	{ O{} l m }{ \mathbb{C}_{#1}\bqty#2{ #3 } }
\DeclareDocumentCommand \tind 		{  }{ k }
\DeclareDocumentCommand \nlf 		{ s }{ \IfBooleanTF{#1}{g}{\vb{g}} }
\DeclareDocumentCommand \dynf 		{  }{ \vb{f} }
\DeclareDocumentCommand \obsf 		{  }{ \vb{h} }
\DeclareDocumentCommand \stVar 		{ s }{ \IfBooleanTF{#1}{x}{\vb{x}} }
\DeclareDocumentCommand \obsVar		{ s }{ \IfBooleanTF{#1}{z}{\vb{z}} }
\DeclareDocumentCommand \stNoise	{ s }{ \IfBooleanTF{#1}{q}{\vb{q}} }
\DeclareDocumentCommand \obsNoise	{ s }{ \IfBooleanTF{#1}{r}{\vb{r}} }
\DeclareDocumentCommand \stNoiseCov	{ s }{ \IfBooleanTF{#1}{\sigma^2_q}{\mb{Q}} }
\DeclareDocumentCommand \obsNoiseCov{ s }{ \IfBooleanTF{#1}{\sigma^2_r}{\mb{R}} }
\DeclareDocumentCommand \stMean		{  }{ \vb{m}^x_{\tind |\tind-1} }
\DeclareDocumentCommand \obsMean	{  }{ \vb{m}^z_{\tind |\tind-1} }
\DeclareDocumentCommand \stCov		{  }{ \mb{P}^x_{\tind |\tind-1} }
\DeclareDocumentCommand \stObsCov	{  }{ \mb{P}^{xz}_{\tind |\tind-1}}
\DeclareDocumentCommand \obsStCov	{  }{ \mb{P}^{zx}_{\tind |\tind-1}}
\DeclareDocumentCommand \obsCov		{  }{ \mb{P}^z_{\tind |\tind-1} }
\DeclareDocumentCommand \stDim		{  }{ {d_x} }
\DeclareDocumentCommand \stNoiseDim	{  }{ {d_q} }
\DeclareDocumentCommand \obsDim		{  }{ {d_z} }
\DeclareDocumentCommand \obsNoiseDim{  }{ {d_r} }
\DeclareDocumentCommand \filtMean 	{ O{\tind} }{ \vb{m}^x_{#1|#1} }
\DeclareDocumentCommand \filtCov 	{ O{\tind} }{ \mb{P}^x_{#1|#1} }
\DeclareDocumentCommand \inVarU		{  }{ \vb{\xi} }
\DeclareDocumentCommand \inVar 		{  }{ \vb{x} }
\DeclareDocumentCommand \inMean 	{  }{ \vb{m} }
\DeclareDocumentCommand \inCov 		{  }{ \mb{P} }
\DeclareDocumentCommand \inCovFct	{  }{ \mb{L} }
\DeclareDocumentCommand \inDim 		{  }{ D }
\DeclareDocumentCommand \outVar 	{ s }{ \IfBooleanTF{#1}{y}{\vb{y}} }
\DeclareDocumentCommand \outMean 	{  }{ \vb{\mu} }
\DeclareDocumentCommand \outCov 	{  }{ \mb{\Pi} }
\DeclareDocumentCommand \outDim 	{  }{ E }
\DeclareDocumentCommand \inoutCov 	{  }{ \mb{C} }

\DeclareDocumentCommand \outMeanApp	{  }{ \hat{\vb{\mu}} }
\DeclareDocumentCommand \outCovApp 	{  }{ \hat{\mb{\Pi}} }
\DeclareDocumentCommand \inoutCovApp{  }{ \hat{\mb{C}} }
\DeclareDocumentCommand \inDof 		{  }{ \nu }
\DeclareDocumentCommand \nlfDof 	{  }{ \nu_{\nlf*} }
\DeclareDocumentCommand \filtDof 	{  }{ \nu^{\star} }
\DeclareDocumentCommand \stDof 		{  }{ \nu_x }
\DeclareDocumentCommand \obsNoiseDof{  }{ \nu_r }
\DeclareDocumentCommand \stNoiseDof	{  }{ \nu_q }
\DeclareDocumentCommand \R 			{  }{ \mathbb{R} }
\DeclareDocumentCommand \D 			{  }{ \mathcal{D} }
\DeclareDocumentCommand \wm 		{  }{ \vb{w} }
\DeclareDocumentCommand \wc 		{  }{ \mb{W} }
\DeclareDocumentCommand \wcc 		{  }{ \mb{W}_c }

\DeclareDocumentCommand \tpExpVar	{  }{ \mb{S} }
\DeclareDocumentCommand \nlfObs		{ s }{ \IfBooleanTF{#1}{\mb{y}}{\vb{y}} }
\DeclareDocumentCommand \kerMean	{  }{ \vb{q} }
\DeclareDocumentCommand \kerCov		{  }{ \mb{Q} }
\DeclareDocumentCommand \kerCCov	{  }{ \mb{R} }
\DeclareDocumentCommand \kerMat		{  }{ \mb{K} }
\DeclareDocumentCommand \kerPar		{  }{ \vb{\theta} }
\DeclareDocumentCommand \kerf		{  }{ k }

\DeclareDocumentCommand \rbfLam		{  }{ \mb{\Lambda} }
\DeclareDocumentCommand \rbfScale   {  }{ s }

\DeclareDocumentCommand \trNum		{  }{ N }
\DeclareMathOperator{\atantwo}{\mathrm{atan2}}

\DeclareUnicodeCharacter{00A0}{ } 

\title{\textbf{\student{} Process Quadratures for Filtering of Non-Linear Systems with Heavy-Tailed Noise}}
\author[1]{Jakub Prüher}
\author[2]{Filip Tronarp}
\author[2]{Toni Karvonen}
\author[2]{Simo Särkkä}
\author[1]{Ond\v{r}ej Straka}

\affil[1]{Department of Cybernetics, University of West Bohemia, Czech Republic}
\affil[2]{Department of Electrical Engineering and Automation, Aalto University, Finland}

\begin{document}

\maketitle

\begin{abstract}
	\noindent The aim of this article is to design a moment transformation for \student{} distributed random variables, which is able to account for the error in the numerically computed mean. 
	We employ \student{} process quadrature, an instance of Bayesian quadrature, which allows us to treat the integral itself as a random variable whose variance provides information about the incurred integration error.
	Advantage of the \student{} process quadrature over the traditional Gaussian process quadrature, is that the integral variance depends also on the function values, allowing for a more robust modelling of the integration error.
	The moment transform is applied in nonlinear sigma-point filtering and evaluated on two numerical examples, where it is shown to outperform the state-of-the-art moment transforms.
\end{abstract}

\section{Introduction}\label{sec:intro}

State estimation problems arise in many engineering fields and sciences, such as global positioning system~\citep{Grewal2007}, tracking~\citep{Blackman1999,Li2013} and finance~\citep{Bhar2010}.
In this paper, we are interested in designing a robust filter applicable in cases where the process and measurement noises are heavy-tailed.
\student{} filter for linear systems was presented by~\citet{Roth2013}, where it was found to increase robustness with respect to assumptions on the noise statistics in the sense of mean error. 
\citet{Tronarp2016} later extended the filter to the non-linear and non-additive noise case by generalising the Unscented transform to integration with respect to a \student{} distribution; the robustness with respect to noise assumptions was also replicated.
The key component of any local filtering algorithm is a moment transform (MT), which is responsible for computing the moments of a random variable transformed through a non-linear function. 
When the dynamical system is linear, the transformed moments can be computed analytically.

In the case of non-linear systems, numerical approximations are required. 
Many such approximations have been proposed over the years, including the well-known Unscented transform~\citep{JulierUhlmann2004} (an instance of a fully symmetric integration rule~\citep{McNamee1967,Tronarp2016}), and the Gauss--Hermite quadrature~\citep{ItoXiong2000,WuHuWuHu2006}.

All the above-mentioned MTs are based on numerical quadrature rules that are approximative and make errors that go unaccounted for. 
Recently,~\citet{Prueher2017} proposed to leverage the Bayesian quadrature (BQ) approach and presented a Gaussian process quadrature (GPQ) MT which is able to account for the integration error in a principled manner.
In particular, Gaussian filters with the GPQ MTs were shown to exhibit better filter self-assessment~\citep{Prueher2017,Li2006} properties than their classical quadrature counterparts.
It is thus reasonable to expect similar improvements in the setting of \student{} filtering~\citep{Huang2016,Tronarp2016}.

The foundation of the GPQ is the Gaussian process (GP) regression model \citep{Rasmussen2006}, which is used as a surrogate probabilistic model of the integrand. 
Given the evaluation points (sigma-points), the mean function of the conditioned GP approximates the integrand, while the GP predictive variance informs about the lack of knowledge of the function behaviour in places where it was not evaluated.
As the numerical quadrature approximations seek to work with as few integrand evaluations as possible to constrain the computational load, it is of paramount importance that the surrogate model predictive variance is as accurate as possible.

An attractive alternative to the GP is the \student{} process (TP) regression~\citep{Shah2014,SolinSarkka2015}, which has the advantage that the model predictive variance also depends on the observed function values unlike in the case of the GP model. 
It can thus provide more accurate and robust predictive variances, which directly translate into the resulting integral quadrature approximations.
Motivated by findings of~\citet{Shah2014}, who concluded that \textquote{TP has many if not all of the benefits of GPs, but with increased modelling flexibility at no extra cost.}, we aim to leverage the TP regression model for the design of a \student{} process quadrature (TPQ) MT.
As far as we are aware, the use of TP in a quadrature context has not been previously attempted, although it has been suggested by~\citet[Section 2.1]{Briol2016}.

It should be noted that the transformation of moments is a more general problem as it arises in other applications, such as sensor system design~\citep{Zangl2008} and optimal control~\citep{Ross2015}; however, the focus of this article is on sigma-point filtering~\citep{Sarkka2013a} context.
We combine the proposed TPQ moment transform with the recent approach by~\citet{Tronarp2016} to sigma-point filtering of non-linear systems with heavy-tailed noise.

The article is structured as follows, \Cref{sec:t-process_quadrature} describes the \student{} process quadrature, which is applied in \Cref{sec:general_tpq_transform} for moment transformation design.
\Cref{sec:filtering} applies the proposed TPQ moment transform in \student{} sigma-point filtering, while \Cref{sec:experiments} presents the numerical experiments.
Finally, \Cref{sec:conclusion} concludes the article.

\section{\student{} Process Quadrature}\label{sec:t-process_quadrature}

The key problem in moment transformations pertains to the computation of the moments of a transformed random variable.
Given a random variable $\inVar \in \R^\inDim$ with a density function $p$ and a non-linear integrand $\nlf \colon \R^\inDim \to \R^\outDim$, the goal of an MT is to compute
\begin{align}
\outMean 	&= \E[\inVar]{\nlf(\inVar)} = \int \nlf(\inVar) p(\inVar)\d[\inVar], \label{eq:mean_integral}\\
\outCov		&= \Cov[\inVar]{\nlf(\inVar)} = \int (\nlf(\inVar) - \outMean)(\nlf(\inVar) - \outMean)\T p(\inVar) \d[\inVar], \label{eq:cov_integral}\\
\inoutCov 	&= \Cov[\inVar]{\inVar,\nlf(\inVar)} = \int (\inVar - \E{\inVar})(\nlf(\inVar) - \outMean)\T p(\inVar)\d[\inVar]. \label{eq:croscov_integral}
\end{align}
In general, the above integrals cannot be evaluated analytically. There are essentially two ways of approximating them.
The first one is to linearize the integrand (in the context of state estimation, this yields the extended Kalman filter and its relatives) and the second to use numerical quadrature which leads to variants of sigma-point filters. 
This section develops a probabilistic numerical integration rule based on modelling the integrand as a \student{} process. 
Its application to the MT problem is given in \Cref{sec:general_tpq_transform}.

A quadrature (or a sigma-point) rule is an approximation to the expectation (integral) of the transformed random variable $\nlf(\inVar)$ of the form
\begin{equation}\label{eq:quadrature}
\E[\inVar]{\nlf(\inVar)} = \int \nlf(\inVar) p(\inVar)\d[\inVar] \approx \sum_{i=1}^N w_i \nlf(\inVar_i)
\end{equation}
where $w_i \in \R$ are the non-zero \emph{weights} and $\inVar_i \in \R^\inDim$ are the \emph{sigma-points} (or \emph{nodes}).
The classical approach to selecting the sigma-points and the weights is to choose them so that the rule~\labelcref{eq:quadrature} is exact whenever the coordinates of $\nlf$ are multivariate polynomials of low degree.
The most popular sigma-point rules used in Gaussian assumed density filtering, such as the Unscented transform~\citep{JulierUhlmann2004}, an instance of a fully symmetric integration rule~\citep{McNamee1967}, and iterated Gauss--Hermite quadrature~\citep{ItoXiong2000}, belong to this category.

Alternatively, one can embrace the philosophy of \emph{probabilistic numerics}~\citep{Diaconis1988,OHagan1992,HennigOsborneGirolami2015,Cockayne2017} and view the process of numerical computation of an integral as a problem of statistical inference.
In this setting, the integrand is modelled as a stochastic process conditioned on the evaluations $\nlf(\inVar_i)$ at the sigma-points $\inVar_i$. 
Injecting additional uncertainty through a stochastic process into the problem might seem counter-productive at first. 
However, after examining \labelcref{eq:quadrature}, we come to realize that the quadrature rule sees the function only through a finite number of function values---how the integrand behaves at other points is unknown.
A probabilistic model for the integrand allows us to acknowledge this uncertainty and induces a posterior probability distribution over the integral~\eqref{eq:quadrature} itself. 
The posterior integral mean estimates the value of the integral while the posterior variance is construed as a model of the integration error.
Numerical approximations of this sort go by the name \emph{Bayesian quadrature} (BQ)~\citep{OHagan1991,Minka2000,Briol2016}.

Gaussian processes~\citep{Rasmussen2006} have been a popular modelling choice in the BQ as prior distributions on functions due to their favourable analytical properties.
Namely, when a GP distributed function is mapped through an integral, which is a \emph{linear} operator, the integration result is also Gaussian distributed.
GPQ has been applied to filtering problems in~\citep{Sarkka2016,Prueher2016} (a more general presentation on the MT problem can be found in~\citep{Prueher2017}).

In this article, we consider the TP regression model as an attractive alternative to the GP, which we believe has potential to bring about significant improvements for the reasons outlined below.
Since \student{} distributions are invariant under affine transformations, TPs retain the favourable analytical properties of GPs, while providing increased modelling flexibility~\citep{Shah2014}, namely:
\begin{itemize}[noitemsep]
	\item the distribution of the integral itself is \student{},
	\item as opposed to GP, the model predictive variance additionally depends on the function values, which allows for more precise uncertainty modelling,
	\item and finally, a TP contains a GP as a special case (for infinite degrees of freedom (DoF)).
\end{itemize}

\subsection{\student{} Process Regression Model}\label{ssec:t-process_regression}

From the perspective of the BQ, the TP regression model is a tool for modelling uncertainty in the knowledge of the numerically integrated function.
Consider a real-valued function $\nlf*\colon\R^\inDim\to\R$ which is assigned a TP prior, such that \( \nlf*(\inVar)~\sim~\mathcal{TP}(0,\, k(\inVar, \inVar'),\, \nlfDof) \).
This implies that for any finite collection of points $\inVar_1',\, \ldots,\, \inVar_m'$ the function values are jointly \student{} distributed with the degrees of freedom (DoF) $\nlfDof~>~2$. That is,
\begin{equation}\label{eq:tp_prior}
\St*[\bmqty{\nlf*(\inVar_1') & \cdots & \nlf*(\inVar_m')}][\vzero][\tfrac{\nlfDof-2}{\nlfDof}\kerMat][\nlfDof],
\end{equation}
where the kernel (covariance) matrix $\kerMat$ is made up of pairwise kernel evaluations, so that $[\kerMat]_{ij} = \kerf(\inVar_i',\inVar_j'; \kerPar)$, where \( \kerPar \) are the kernel parameters.
For brevity, dependence on \( \kerPar \) will be made explicit only when absolutely necessary.
The choice of the positive-definite kernel $\kerf$ is up to the user and usually reflects expected smoothness of the underlying function.
Given $\trNum$ observations $\nlfObs = [\nlf*(\inVar_1),\ldots,\nlf*(\inVar_\trNum)]\T$ at the evaluation points $\inVar_1,\, \ldots,\, \inVar_\trNum$, conditioning on the data $\D = \{ (\inVar_i,\, \nlf*(\inVar_i)) \}_{i=1}^\trNum$ results in a TP posterior with mean and variance~\citep{Shah2014,SolinSarkka2015}
\begin{align}
	\E[\nlf*]{\nlf*(\inVar) \mid \D} &= \vb{\kerf}\T(\inVar)\kerMat\I\nlfObs, \label{eq:tp-regression_mean} \\
	\V[\nlf*]{\nlf*(\inVar) \mid \D} &= \frac{\nlfDof - 2 + \nlfObs\T\kerMat\I\nlfObs}{\nlfDof - 2 + \trNum }\bqty{\kerf(\inVar, \inVar) - \vb{\kerf}\T(\inVar)\kerMat\I\vb{\kerf}(\inVar)}, \label{eq:tp-regression_variance}
\end{align}
where \( \bqty{\vb{k}(\inVar)}_{i} = \kerf(\inVar, \inVar_i) \). 

\begin{figure}[!htb]
	\centering
	\input{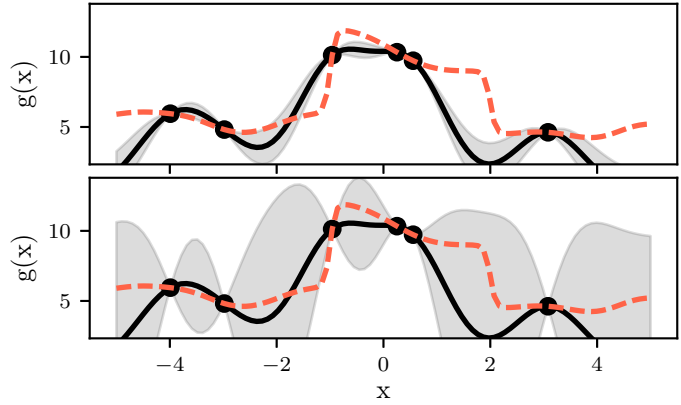}
	\caption{Comparison of predictive moments of the Gaussian process (top) and the \student{} process (bottom) regression models using the same set of kernel parameters. The DoF of the TP model was set to \( \nlfDof = 10 \). The true function (dashed red), the posterior mean (solid black) and predictive variance (gray band).}
	\label{fig:gp_vs_tp}
\end{figure}

The posterior mean is identical to that of the GP regression, but the posterior variance~\labelcref{eq:tp-regression_variance} has the additional data-dependent scaling coefficient $(\nlfDof - 2 + \nlfObs\T\kerMat\I\nlfObs)/(\nlfDof - 2 + \trNum)$. 
This dependency on the function evaluation means that the TP regression is often more informative about the true underlying function than GP regression. 
Furthermore, the DoF is an additional tunable parameter allowing for control of the heavy-tailed process behaviour.
The lower the DoF, the heavier the tails and vice versa.
For \( \nlfDof \to \infty \) a GP is recovered, which means the GP regression can be interpreted as a special case of the TP regression.
Worth noting is that for increasing DoF the scaling factor becomes less dependent on the function values, eventually degrading to the GP predictive variance.
Intuitively, one would assume that the predictive variance of the function would be affected by its own observations.
For the GP this is not the case, implying that it is possible to know the predictive variance before the function observations are even obtained, which can be advantageous in certain applications.
We argue, however, that in BQ applications, where limited datasets are encountered and an accurate quantification of uncertainty is crucial, the TP should be preferred, because of its superior characterization of uncertainty. 
The difference in the predictive variance can be seen in \Cref{fig:gp_vs_tp}, where the GP and TP are compared using the same values of kernel parameters.
The mean functions of both models, which approximate the true underlying function,  are identical.
The TP is able to inflate the predictive variance due to its heavy-tailed nature, resulting in more realistic uncertainties in the function behaviour given the available data.

\subsection{\student{} Process Quadrature Integral Moments}\label{ssec:integral_moments}

A TP posterior distibution over the integrand \( \nlf*(\inVar) \) induces a posterior distribution over the integral $\E[\inVar]{\nlf*(\inVar)} = \int \nlf*(\inVar) p(\inVar)\d[\inVar]$. 
The mean and variance of this distribution can be directly evaluated~\citep{Minka2000,Briol2016}:
\begin{align}
	\E[\nlf*]{\E[\inVar]{\nlf*(\inVar)} \mid \D} 
	&= \E[\inVar]{\E[\nlf*]{\nlf*(\inVar) \mid \D}} = \kerMean\T\kerMat\I\nlfObs, \label{eq:tpq_integral_mean} \\
	\V[\nlf*]{\E[\inVar]{\nlf*(\inVar)} \mid \D} 
	&= \E[\inVar, \inVar']{\Cov[\nlf*]{\nlf*(\inVar), \nlf*(\inVar') \mid \D}} = \gamma\bqty{ \E[\inVar,\inVar']\big{\kerf(\inVar, \inVar')} - \kerMean\T\kerMat\I\kerMean }, \label{eq:tpq_integral_variance}
\end{align}
where \( \gamma =  (\nlfDof - 2 + \nlfObs\T\kerMat\I\nlfObs)/(\nlfDof - 2 + \trNum) \) and \( \kerMean = \E[\inVar]\big{\vb{\kerf}(\inVar)}\).
In \Cref{eq:tpq_integral_mean}, the mean of the integral is identical to integrating the TP posterior mean function.
The only difference to the GPQ is in the integral posterior variance \labelcref{eq:tpq_integral_variance} having a data-dependent coefficient.
Note that the integral posterior mean~\eqref{eq:tpq_integral_mean} is indeed a quadrature rule of the form~\eqref{eq:quadrature} with the weights $w_i = [\kerMean\T\kerMat\I]_i$.
Integrals of general vector functions $\nlf\colon\R^\inDim\to\R^\outDim$ can be evaluated by applying the above equations to each function output independently with the same kernel and DoF used for each output.

\section{\student{} Process Quadrature Moment Transform}\label{sec:general_tpq_transform}

In this section, we apply the results for the integral moments from the previous section for derivation of our proposed general purpose TPQ moment transform. 
The results from this section are aimed to be applied later in \Cref{sec:filtering} for approximation of the state and measurement moments in a \student{} sigma-point filtering setting. 

First, consider a general case of non-linear transformation of arbitrarily distributed random variable
\begin{equation}\label{eq:general_non-linear_transformation}
	\outVar = \nlf(\inVar), \qquad \inVar\sim p(\inVar)
\end{equation}
where \( \nlf:\R^\inDim\to\R^\outDim \) is such that the first two moments of \( \outVar \) exist.
In the BQ, when the uncertainty in $\nlf(\inVar)$ is introduced through a posterior stochastic process regression model, the general BQ approximations to the moments in \Crefrange{eq:mean_integral}{eq:croscov_integral} need to account for it, resulting in 
\begin{align}
	\E{\outVar} = \E[\inVar]{\nlf(\inVar)} &\approx \E[\nlf, \inVar]{\nlf(\inVar)}, \\
	\Cov{\outVar} = \Cov[\inVar]{\nlf(\inVar)} &\approx \Cov[\nlf, \inVar]{\nlf(\inVar)}, \\
	\Cov{\inVar, \outVar} = \Cov{\inVar,\, \nlf(\inVar)} &\approx \Cov[\nlf, \inVar]{\inVar,\, \nlf(\inVar)}.
\end{align}
The BQ transformed mean can be written out, using the law of total expectation, as
\begin{equation}\label{eq:bq_mean_decomposition}
	\E[\nlf, \inVar]{\nlf(\inVar)} = \E[\nlf]{\E[\inVar]{\nlf(\inVar)}} = \E[\inVar]{\E[\nlf]{\nlf(\inVar)}},
\end{equation}
which shows that the expectation of the integral is equivalent to integrating the model mean function (cf.\ \Cref{eq:tpq_integral_mean}).
Applying the law of total covariance, we can decompose the transformed covariance in two ways
\begin{align}\label{eq:bq_variance_decomposition}
	\Cov[\nlf, \inVar]{\nlf(\inVar)} 
    &= \E[\nlf]{\Cov[\inVar]{\nlf(\inVar)}} + \Cov[\nlf]{\E[\inVar]{\nlf(\inVar)}} \\
    &= \E[\inVar]{\Cov[\nlf]{\nlf(\inVar)}} + \Cov[\inVar]{\E[\nlf]{\nlf(\inVar)}}.
\end{align}
The first decomposition reveals the fact that the general BQ MT incorporates integral variance as an additional term in transformed variance.
By now it is evident that the MT based on BQ can be interpreted as a principled covariance inflation scheme.
Note that for deterministic \( \nlf \) the covariance decomposition in \Cref{eq:bq_variance_decomposition} would revert back to the classical definition of the covariance. 
In the BQ, however, the integrand uncertainty is crucial for modelling the integration error.
The second decomposition is computationally beneficial because the individual terms can be computed in closed form for suitable kernel-density pairs. 

In order to derive the TPQ MT for \( \St*[\inVar][\inMean][\inCov][\inDof] \), we utilize the familiar stochastic decoupling substitution \( \inVar = \inMean + \inCovFct\inVarU \), which allows for casting the expectations in terms of a standard \student{} random variable \( \St*[\inVarU][\vzero][\eye][\inDof] \), so that
\begin{align}\label{}
    \E[\inVar]{\nlf(\inVar)} &= \E[\inVarU]{\nlf(\inMean + \inCovFct\inVarU)} \approx \E[\nlf, \inVarU]{\nlf(\inMean + \inCovFct\inVarU)}, \\
    \Cov[\inVar]{\nlf(\inVar)} &= \Cov[\inVarU]{\nlf(\inMean + \inCovFct\inVarU)} \approx \Cov[\nlf, \inVarU]{\nlf(\inMean + \inCovFct\inVarU)}, \\
    \Cov[\inVar]{\inVar, \nlf(\inVar)} &= \Cov[\inVarU]{\inVarU, \nlf(\inMean + \inCovFct\inVarU)} \approx \Cov[\nlf, \inVarU]{\inVarU, \nlf(\inMean + \inCovFct\inVarU)},
\end{align}
where \( \inCovFct\inCovFct\T = \inCov \).
For notational brevity we define \( \widetilde{\nlf}(\inVarU) \triangleq \nlf(\inMean + \inCovFct\inVarU) \).
Using \Cref{eq:tpq_integral_mean,eq:bq_mean_decomposition}, the transformed mean becomes
\begin{equation}\label{}
	\E[\nlf, \inVarU]{\widetilde{\nlf}(\inVarU)} = \E[\inVarU]{\E[\nlf]{\widetilde{\nlf}(\inVarU)}} = \nlfObs*\T\kerMat\I\E[\inVarU]\big{\vb{\kerf}(\inVarU)},
\end{equation}
where \( \nlfObs* = \bmqty{ \nlfObs_1 & \ldots & \nlfObs_\outDim} \) contains observations of each output of \( \nlf \) in columns.
For the transformed covariance we employ \Cref{eq:bq_variance_decomposition} and obtain
\begin{equation}\label{}
	\Cov[\nlf, \inVarU]{\widetilde{\nlf}(\inVarU)} = \E[\inVarU]\big{\E[\nlf]\big{\widetilde{\nlf}(\inVarU)}\E[\nlf]\big{\widetilde{\nlf}(\inVarU)}\T} - \outMeanApp\outMeanApp\T + \E[\inVarU]{\Cov[\nlf]{\widetilde{\nlf}(\inVarU)}},
\end{equation}
where, after plugging in from \Cref{eq:tpq_integral_mean}, the first term becomes
\begin{equation}
\E[\inVarU]\big{\E[\nlf]\big{\widetilde{\nlf}(\inVarU)}\E[\nlf]\big{\widetilde{\nlf}(\inVarU)}\T} = \nlfObs*\T\kerMat\I\E[\inVarU]\big{\vb{\kerf}(\inVarU)\vb{\kerf}(\inVarU)\T}\kerMat\I\nlfObs*
\end{equation}
and the third term is
\begin{equation}
\E[\inVarU]{\Cov[\nlf]{\widetilde{\nlf}(\inVarU)}} = \diag{\bmqty{s^1_{\mathrm{TP}} & \ldots & s^\outDim_{\mathrm{TP}}}}
\end{equation}
where
\begin{equation}\label{eq:tpq_expected_model_covariance}
	s^e_{\mathrm{TP}} = \gamma_e\bqty{\E[\inVarU]{\kerf(\inVarU, \inVarU)} - \trace\pqty\big{\E[\inVarU]\big{\vb{\kerf}(\inVarU)\vb{\kerf}(\inVarU)\T}\kerMat\I}}
\end{equation}
and \( \gamma_e = (\nlfDof - 2 + \nlfObs_e\T\kerMat\I\nlfObs_e)/(\nlfDof - 2 + \trNum) \).
Finally, for the input-output covariance we have
\begin{equation}\label{}
	\Cov[\nlf, \inVarU]{\inVarU, \widetilde{\nlf}(\inVarU)} = \inCovFct\E[\inVarU]{\inVarU\E[\nlf]{\widetilde{\nlf}(\inVarU)}} = \inCovFct\E[\inVarU]\big{\inVarU\vb{\kerf}(\inVarU)\T}\kerMat\I\nlfObs*.
\end{equation}

The following definition gathers our results so far.
\begin{defn}[General TPQ moment transform]\label{def:tpq_moment_transform}
	The general \student{} process quadrature approximation to the joint distribution of  $ \St*[\inVar][\inMean][\frac{\nu-2}{\nu}\inCov] $ and a transformed random variable $ \outVar = \nlf(\inVar) $ is given by
	\begin{equation}
	\St*[\bmqty{\inVar \\ \outVar}][\bmqty{\inMean \\ \outMeanApp}][\frac{\nu-2}{\nu}\bmqty{\inCov & \inoutCovApp \\ \inoutCovApp\T & \outCovApp}][\inDof]
	\end{equation}
	where the transformed moments are
	\begin{align}
	\outMeanApp 	&= \nlfObs*\T\wm, \label{eq:tpq_mean_out} \\
	\outCovApp 		&= \nlfObs*\T\wc\nlfObs* - \outMeanApp\outMeanApp\T + \tpExpVar, \label{eq:tpq_cov_out} \\
	\inoutCovApp 	&= \inCovFct\wcc\nlfObs*, \label{eq:tpq_covio_out} \\
	\tpExpVar 		&= \diag{\bmqty{s^1_{\mathrm{TP}} & \ldots & s^\outDim_{\mathrm{TP}}}}, \\
	s^e_{\mathrm{TP}} &= \frac{\nlfDof - 2 + \nlfObs_e\T\kerMat\I\nlfObs_e}{\nlfDof - 2 + \trNum}\bqty{\bar{k} - \trace\pqty\big{\kerCov\kerMat\I}} \label{eq:tpq_expected_gp_variance}
	\end{align}
	and $ \nlfObs* = \bmqty{\nlfObs_1 & \ldots & \nlfObs_\outDim} $, where the $ e $-th column \( \nlfObs_e = \bmqty{\nlf*_e(\inVar_1) & \ldots & \nlf*_e(\inVar_\trNum)}\T \) contains function values of the $ e $-th output of $ \nlf(\inVar) $.
	The sigma-points are given by \( \inVar_i = \inMean + \inCovFct\inVarU_i \), where \( \inCovFct\inCovFct\T = \inCov \).
	The elements of the kernel matrix are $ \bqty{\kerMat}_{ij} = \kerf(\inVarU_i, \inVarU_j;\kerPar) $ and the TPQ weights are $ \wm = \kerMat\I\kerMean $, $ \wc = \kerMat\I\kerCov\kerMat\I $ and $ \wcc = \kerCCov\kerMat\I $, where
	\begin{align}
	\bqty{\kerMean}_{i} 	&= \E[\inVarU]{\kerf\pqty{\inVarU, \inVarU_i; \kerPar}}, \label{eq:tpq_kernel_mean} \\
	\bqty{\kerCov}_{ij} 	&= \E[\inVarU]{\kerf\pqty{\inVarU, \inVarU_i; \kerPar}\kerf\pqty{\inVarU, \inVarU_j; \kerPar}}, \label{eq:tpq_kernel_covariance} \\
	\bqty{\kerCCov}_{*j} 	&= \E[\inVarU]{\inVar\kerf\pqty{\inVarU, \inVarU_j; \kerPar}}, \label{eq:tpq_kernel_crosscovariance} \\
	\bar{k}					&= \E[\inVarU]{\kerf\pqty{\inVarU, \inVarU; \kerPar}}. \label{eq:tpq_kernel_exp}
	\end{align}
	The notation $\bqty{\kerCCov}_{*j}$ stands for the $j$th column of the matrix $\kerCCov$.
	The set of unit sigma-points \( \Bqty{\inVarU_i:\ i = 1, \ldots, \trNum} \) can be chosen \emph{arbitrarily}.
\end{defn}
The transform is general in a sense that it can, in principle, operate with any kernel.
Since decoupling is used in the moment integrals, the kernel expectations in \labelcref{eq:tpq_kernel_mean}--\labelcref{eq:tpq_kernel_exp} do not depend on the parameters of the distribution of $\inVar$ and the TPQ weights can be fully pre-computed.
This fact significantly eases computational burden when our proposed TPQ moment transform is later applied in \student{} sigma-point filtering.

\subsection{Kernel Expectations}\label{ssec:kernel_expectations}

From the above summary it is apparent that the TPQ MT requires evaluation of the kernel expectations in \labelcref{eq:tpq_kernel_mean}--\labelcref{eq:tpq_kernel_exp}.
The popular radial basis function (RBF) kernel, given by
\begin{equation}\label{eq:rbf_kernel}
	\kerf(\inVarU, \inVarU'; \kerPar) = \rbfScale^2\exp\pqty{-\frac{1}{2}\pqty{\inVarU - \inVarU'}\T\rbfLam\I\pqty{\inVarU - \inVarU'}},
\end{equation}
where \(\rbfScale,\ \rbfLam = \diag{\bmqty{\ell^2_1 & \ldots & \ell^2_\inDim}} \) are kernel parameters collected into a vector \( \kerPar \), has been used in our previous work \citep{Prueher2016}. 
This kernel admits a closed-form evaluation of the expectations in \labelcref{eq:tpq_kernel_mean}--\labelcref{eq:tpq_kernel_exp} for a Gaussian distributed $\inVar$.
However, in our case, where \( \inVar \) is \student{} distributed, we have been unable to find any kernel admitting closed-form solution.
For this reason, we used the RBF kernel and resorted to the standard Monte Carlo numerical approximation.
Fortunately, in filtering this is not really a problem, because the expectations, and consequently the TPQ weights, can be fully pre-computed offline.

\section{TPQ \student{} Filter}\label{sec:filtering}
The aim of this section is to present the TPQ \student{} filter harnessing the proposed TPQ MT presented in \Cref{sec:general_tpq_transform} and enriching the work of~\citet{Tronarp2016} with the BQ philosophy. 
We first review the general \student{} filtering framework for non-additive noise setting, and then later apply the TPQ MT for construction of the TPQ \student{} filter.

Consider the following discrete-time state-space model
\begin{align}
	\stVar_\tind 	&= \dynf(\stVar_{\tind-1},\stNoise_{\tind-1}), 	\label{eq:ssm_dynamics} \\
	\obsVar_\tind	&= \obsf(\stVar_\tind,\obsNoise_\tind),			\label{eq:ssm_measurement_model}
\end{align}
where \labelcref{eq:ssm_dynamics} describes the evolution of the system state \( \stVar_{\tind} \) in time and \labelcref{eq:ssm_measurement_model} describes the process by which measurements \( \obsVar_\tind \) are generated.
The function $\dynf\colon \R^\stDim \times \R^\stNoiseDim \to \R^\stDim$ is the system dynamics and $\obsf \colon \R^\stDim \times \R^{\obsNoiseDim} \to \R^\obsDim$ is the measurement function.
The variables $\stNoise_{\tind}$ and $\obsNoise_{\tind}$ represent the zero-mean process and measurement noises with known covariance matrices $\stNoiseCov_{\tind}$ and $\obsNoiseCov_\tind$, respectively.  

The source of novelty in \student{} filter comes from the conditioning formula for \student{} random variables~\citep{BarndorffNielsenKentSorenson1982,Roth2013}.
The measurement update equations can be derived by assuming that the state and the measurement are jointly \student{} distributed, such that
\begin{equation}\label{eq:joint_t_approximation}
	\St*[\bmqty{\stVar_\tind \\ \obsVar_\tind }][\bmqty{\stMean \\ \obsMean}][\frac{\inDof -  2}{\inDof }\bmqty{\stCov & \stObsCov \\ \obsStCov & \obsCov}][\inDof].
\end{equation}
Then, the conditioned state is distributed according to \( \stVar_\tind~\mid~\obsVar_{1:\tind} \sim \mathrm{St}\pqty\big{\filtMean,\, \tfrac{\filtDof - 2}{\filtDof}\filtCov,\, \filtDof} \) with the statistics given by
\begin{align}
	\filtMean 	&= \stMean + \stObsCov(\obsCov)\I(\obsVar_\tind - \obsMean), \label{eq:student_update_mean}\\
	\filtCov	&= \frac{\stDof - 2 + \beta }{\stDof - 2 + \obsDim} \pqty{\stCov - \stObsCov(\obsCov)\I\obsStCov}, \label{eq:student_update_cov}\\
	\beta 		&= (\obsVar_\tind - \obsMean)\T(\obsCov)\I(\obsVar_\tind - \obsMean), \\
	\filtDof	&= \inDof + \obsDim. \label{eq:student_update_dof}
\end{align}

These equations constitute the \student{} measurement update rule, where \( \filtMean \) and \( \filtCov  \) are the filtered state mean and covariance, respectively.
It is instructive to consider the case where $\obsVar_\tind$ is Gaussian, which entails $\beta \sim \chi^2(\obsDim)$ and $\E{\beta} =  \obsDim$. 
Thus the posterior covariance $\filtCov$ either increases or decreases depending on the outcome of $\beta$ in relation to its expected value under a $\chi^2$ assumption.

The means and covariances in \labelcref{eq:joint_t_approximation}, necessary for the update, can be approximated by \emph{any} moment transform.
The predicted state mean \( \stMean \) and covariance \( \stCov \) are computed by using the system dynamics \( \dynf(\stVar_{\tind-1},\stNoise_{\tind-1}) \) in place of the general non-linear transformation from \Cref{eq:general_non-linear_transformation} with the input variable distributed according to 
\begin{equation}\label{eq:joint_state_noise}
	\St*[\bmqty{\stVar_{\tind-1} \\ \stNoise_{\tind-1} }][\bmqty{\filtMean[\tind-1] \\ 0}][\frac{\inDof -  2}{\inDof }\bmqty{\filtCov[\tind-1] & 0 \\ 0 & \stNoiseCov}][\inDof].
\end{equation}
The same applies for the measurement mean \( \obsMean \), covariance \( \obsCov \) and cross-covariance \( \stObsCov \), which are computed by using the measurement function \( \obsf(\stVar_\tind, \obsNoise_\tind) \) in place of the non-linear transformation from \Cref{eq:general_non-linear_transformation} with input variable distributed according to 
\begin{equation}\label{eq:joint_measurement_noise}
	\St*[\bmqty{\stVar_{\tind} \\ \obsNoise_\tind }][\bmqty{\stMean \\ 0}][\frac{\inDof -  2}{\inDof }\bmqty{\stCov & 0 \\ 0 & \obsNoiseCov}][\inDof].
\end{equation}

Once the approximations to the predictive state and measurement moments are available, a joint \student{} approximation in \Cref{eq:joint_t_approximation} can be formed with the desired degrees of freedom $\inDof$.
This subsequently enables the use of the update rule given by \Crefrange{eq:student_update_mean}{eq:student_update_dof}.

Notice, the DoF update in \Cref{eq:student_update_dof} entails that \( \filtDof\to\infty \) for \( \tind\to\infty \), which means the filter will asymptotically behave as a Kalman filter.
In order to continue to operate at desired DoF $\inDof$, we opt for the moment matching perspective used in~\citep{Roth2013,Tronarp2016} and assume that the conditioned state \( \stVar_{\tind}~\mid~\obsVar_{1:\tind}\sim\mathrm{St}\pqty\big{\filtMean,\, \tfrac{\inDof-2}{\inDof}\filtCov,\, \inDof} \).

The entire procedure of is outlined in \Cref{alg:tpq_filter}. Note that any sigma-points can be used for the TPQ MT in this algorithm.
\begin{alg}[One step of the TPQ \student{} filter]\label{alg:tpq_filter}
	\ \\
	\emph{Input:} filtered mean \( \filtMean[\tind-1] \), filtered covariance \( \filtCov[\tind-1] \), desired DoF \( \inDof \)\\ 
	\emph{Output:} filtered mean \( \filtMean \) and covariance \( \filtCov \)
	\begin{enumerate}
		\item Use the TPQ MT with kernel parameters \( \kerPar_f \) to compute the predictive state moments $\stMean$ and $\stCov$ using \labelcref{eq:tpq_mean_out,eq:tpq_cov_out} assuming the input is distributed according \labelcref{eq:joint_state_noise}.
		\item Use TPQ MT with kernel parameters \( \kerPar_h \) to compute the predictive measurement moments $\obsMean$, $\obsCov$ and $\stObsCov$ using \labelcref{eq:tpq_mean_out,eq:tpq_cov_out,eq:tpq_covio_out} assuming the input is distributed according \labelcref{eq:joint_measurement_noise}.
		\item Use the \student{} measurement update in \labelcref{eq:student_update_mean,eq:student_update_cov} to compute the filtered mean  \( \filtMean \) and covariance  \( \filtCov \).
		\item Fix the DoF by putting \(\mathrm{St}\pqty\big{\stVar_\tind \mid \filtMean,\, \frac{\inDof-2}{\inDof}\filtCov,\, \inDof} \approx \mathrm{St}\pqty\big{\stVar_\tind \mid \filtMean,\, \frac{\filtDof-2}{\filtDof}\filtCov,\, \filtDof} \).
	\end{enumerate}
\end{alg}

\section{Experimental Results}\label{sec:experiments}
In this section, we compare the performance of the proposed TPQ-based \student{} filters and the \student{} filter introduced by~\citet{Tronarp2016}, which is based on classical quadrature.
In all experiments, we are measuring the filter error by the root mean square error (RMSE)
\begin{equation}\label{eq:rmse}
	\mathrm{RMSE} = \pqty{\frac{1}{K}\sum\limits_{\tind=1}^{K} \|\stVar_{\tind} - \filtMean\|^2}^{1/2}.
\end{equation}
Since the BQ-based MTs, such as TPQ or GPQ, are primarily focused on incorporating additional uncertainty by inflating the estimated covariance, we used the inclination indicator (INC) \citep{Li2006} as a metric which takes into account the estimated state covariance.
The indicator is given by
\begin{equation}\label{eq:inc}
	\mathrm{INC} = \frac{10}{K} \sum_{k=1}^{K} \log_{10}\frac{ \pqty\big{\stVar_\tind - \filtMean}\T \pqty\big{\filtCov}\I \pqty\big{\stVar_\tind - \filtMean} }{ \pqty\big{\inVar_\tind - \filtMean}\T \mb{\Sigma}\I_{k} \pqty\big{\stVar_\tind - \filtMean} },
\end{equation}
where $ \mb{\Sigma}_{k} $ is the sample mean-square-error (MSE) matrix, which can be computed from samples of the true system state trajectories.
When the indicator is \( \mathrm{INC} = 0 \) the estimator is said to be \emph{balanced}, which is to say that the estimated covariance is on average equal to the true state MSE matrix.
For \( \mathrm{INC} > 0 \) the estimator is said to be \emph{optimistic} while for \( \mathrm{INC} < 0 \) it is considered \emph{pessimistic}.

\subsection{Univariate Non-Stationary Growth Model}\label{ssec:ungm}
In the first numerical illustration, we consider the univariate non-stationary growth model (UNGM), which is often used for benchmarking purposes \citep{Gordon1993,Kitagawa1996}.
The system is given by the following set of equations
\begin{align}
	\stVar*_{\tind}  &= 0.5\stVar*_{\tind-1} + \frac{25\stVar*_{\tind-1}}{1 + \stVar*_{\tind-1}^2} + 8\cos(1.2\tind) + \stNoise*_{\tind-1}, \label{eq:ungm_dyn} \\
	\obsVar*_{\tind} &= 0.05\stVar*_{\tind}^2 + \obsNoise*_\tind. \label{eq:ungm_obs}
\end{align}
The initial conditions were drawn from \( \N*[\stVar*_0][0][1] \).
Outliers in the state noise \( \stNoise*_\tind \) and measurement noise \( \obsNoise*_\tind \) were simulated with Gaussian mixtures, such that \( \stNoise*_{\tind} \sim 0.8\mathrm{N}\pqty{0,\, \stNoiseCov*} + 0.2\mathrm{N}\pqty{0,\, 10\stNoiseCov*} \) and \( \obsNoise*_{\tind} \sim 0.8\mathrm{N}\pqty{0,\, \obsNoiseCov*} + 0.2\mathrm{N}\pqty{0,\, 100\obsNoiseCov*} \), where \( \stNoiseCov* = 10 \) and \( \obsNoiseCov* = 0.01 \).
We simulated 500 trajectories for 250 time steps, which were used for evaluation of the RMSE and INC performance metrics.
\begin{table}[!htb]
	\centering
	\begin{tabular}{lrrrr}
		\toprule
		{} 						&  RMSE 		&  STD 		&  INC 		&  STD \\
		\midrule
		UKF             		&     8.6924 	&          0.1517 &    3.0012 &         0.1539 \\
		SF             			&    17.4461 	&          0.6236 &   51.8733 &         0.4417 \\
		TPQSF\((\nlfDof=3)\)  	&     7.5683 	&          0.1091 &    1.5837 &         0.1561 \\
		TPQSF\((\nlfDof=4) \)   &     6.8323 	&          0.1044 &    2.3384 &         0.1713 \\
		TPQSF\((\nlfDof=10) \)  &     6.1423 	&          0.0154 &    5.6910 &         0.0324 \\
		TPQSF\((\nlfDof=100)\)  &	  7.4399  	&    	   0.1550 &   12.5120 &         0.1676 \\
		TPQSF\((\nlfDof=500)\)  &	  7.5709  	&          0.1546 &   13.2104 &         0.1623 \\
		GPQSF           		&     7.6766 	&          0.1554 &   13.5926 &         0.1595 \\
		\bottomrule
	\end{tabular}
	\caption{Performance of TPQSF compared in terms of average RMSE and INC. Standard deviations of the criteria were estimated by bootstrapping. 
		For increasing DOF parameter \( \nlfDof \) of the TP regression model the performance approaches that of the GPQSF.}
	\label{tab:ungm_rmse_inc}
\end{table}

All tested filters used a state-space model with an initial condition distributed according to \( \St*[\stVar*_0][0][\tfrac{\inDof-2}{\inDof}1][\inDof] \) and the following noise statistics \( \St*[\stNoise*_\tind][0][\tfrac{\inDof-2}{\inDof}\stNoiseCov*][\inDof],\ \St*[\obsNoise*_\tind][0][\tfrac{\inDof-2}{\inDof}\stNoiseCov*][\inDof] \), where \( \inDof = 4 \).

We compared the RMSE and INC of our proposed TPQSF with the SF~\citep{Tronarp2016}, the UKF~\citep{JulierUhlmann2004} and a \student{} filter using the GPQ MT~\citep{Prueher2017} (abbreviated GPQSF). 
The TPQSF and the GPQSF will be collectively referred to as the BQ filters.
\student{} filters used the same 3rd degree fully symmetric sigma-point set with \( \kappa=0 \) and the filter DoF fixed at \( \inDof = 4 \).
The kernel parameters for all BQ filters were set to \( \kerPar_f = \bmqty{3 & 1} \) and \( \kerPar_h = \bmqty{3 & 3} \).
\Cref{tab:ungm_rmse_inc} reports MC simulation averages of both metrics along with bootstrapped variances~\citep{Wasserman2007} of the averages (using 10,000 samples).
It is evident that the TPQSFs can outperform all the classical filters (UKF, SF) as well as the GPQSF in terms of both metrics.
The values of INC, being closer to zero, indicate increased estimate credibility.
For increasing DoF of the \student{} process model, we observe the performance of TPQSFs approaching that of GPQSF, which is an expected behaviour, since TPQSF with \( \nlfDof = \infty \) is equivalent to GPQSF.

\subsection{Radar Tracking with Glint Noise}\label{ssec:radar_tracking}
As a second illustration, we consider tracking of a moving object where the range and bearing measurements are corrupted with glint noise.
We adopt the example from \citet{Arasaratnam2007}, where the tracking scenario is described by the following state-space model 
\begin{align}
	\stVar_{\tind} &= \bmqty{1 & \tau & 0 & 0 \\
							 0 & 1 & 0 & 0 \\
						     0 & 0 & 1 & \tau \\
					         0 & 0 & 0 & 1}\stVar_{\tind-1} + \bmqty{\tau^2/2 & 0\\\tau & 0\\0 & \tau^2/2\\0 & \tau}\stNoise_{\tind-1} \label{eq:cv_radar_dynamics} \\
	\obsVar_{\tind} &= \bmqty{\sqrt{x^2_\tind + y^2_\tind} \\ \atantwo\pqty{y_\tind, x_\tind}} + \obsNoise_{\tind} \label{eq:cv_radar_measurement_model}
\end{align}
with the system state being defined as \( \stVar_{\tind} = \bmqty{x_\tind & \dot{x}_\tind & y_\tind & \dot{y}_\tind} \).
The state components \( x_\tind \) and \( y_\tind \) are the Cartesian coordinates of the moving object and the pair \( \dot{x}_\tind,\ \dot{y}_\tind \) stands for the velocity in the respective directions.
During simulations, the discretization interval was \( \tau = \SI{0.5}{s} \), the initial state was drawn from \( \N*[\stVar_0][\vb{m}_0][\mb{P}_0] \) with 
\begin{align}
	\vb{m}_0 &= \bmqty{ \SI{10000}{m} & \SI{300}{ms^{-1}} & \SI{1000}{m} & \SI{-40}{ms^{-1}} }, \\
	\mb{P}_0 &= \diag{\bmqty{ \SI{10000}{m^2} & \SI{100}{m^2s^{-2}} & \SI{10000}{m^2} & \SI{100}{m^2s^{-2}} }}.
\end{align}
The state noise is Gaussian distributed, such that \( \N*[\stNoise_\tind][\vzero][\stNoiseCov] \) with covariance \( \stNoiseCov = \diag{\bmqty{ \SI{50}{m^2s^{-4}} & \SI{5}{m^2s^{-4}} }} \).
The glint noise in the measurements is modelled by a Gaussian mixture 
\begin{equation}\label{eq:glint_noise_mixture}
	\obsNoise_\tind \sim (1-\beta)\mathrm{N}\pqty{\vzero,\, \obsNoiseCov_1} + \beta\mathrm{N}\pqty{\vzero,\, \obsNoiseCov_2}
\end{equation} 
with \( \obsNoiseCov_1 = \diag{\bmqty{ \SI{50}{m^2} & \SI{0.4}{mrad^2} }} \) and \( \obsNoiseCov_2 = \diag{\bmqty{ \SI{5000}{m^2} & \SI{16}{mrad^2} }} \), where \( \beta \) is the glint noise probability.

As in the UNGM experiment, we compared the performance of our proposed TPQSF with the SF, the standard UKF and the GPQSF.
The UKF used \( \kappa = 0 \), following the usual heuristic recommendation.
For the TPQSF we considered two settings of the TP model DoF parameter, \( \nlfDof = 2.2 \) and \( \nlfDof=4 \).
All of the \student{} filters assumed that the initial state, the state noise and the measurement noise were characterized by the \student{} distribution, such that 
\begin{align}\label{}
\stVar_0 		\sim\,& \mathrm{St}\pqty\big{\filtMean[0],\, \tfrac{\stDof-2}{\stDof}\filtCov[0],\, \stDof}, \\
\stNoise_\tind 	\sim\,& \mathrm{St}\pqty\big{\vzero,\, \tfrac{\stNoiseDof-2}{\stNoiseDof}\stNoiseCov,\, \stNoiseDof}, \\
\obsNoise_\tind	\sim\,& \mathrm{St}\pqty\big{\vzero,\, \tfrac{\obsNoiseDof-2}{\obsNoiseDof}\obsNoiseCov,\, \obsNoiseDof}
\end{align}
where the initial state estimate was \( \filtMean[0] = \bmqty{ \SI{10175}{m} & \SI{295}{ms^{-1}} & \SI{980}{m} & \SI{-35}{ms^{-1}}} \), the initial covariance \( \filtCov[0] = \mb{P}_0 \) and DoF parameters were \( \stDof = 1000 \), \( \stNoiseDof = 1000\) and \(\obsNoiseDof = 4.0 \).
The kernel parameters for the BQ filters were set to \( \kerPar_f = \bmqty{1 & 100 & 100 & 100 & 100}  \) for the dynamics model and \( \kerPar_f = \bmqty{0.05 & 10 & 100 & 10 & 100} \) for the measurement model.

\begin{figure}[!htb]
	\centering
	\input{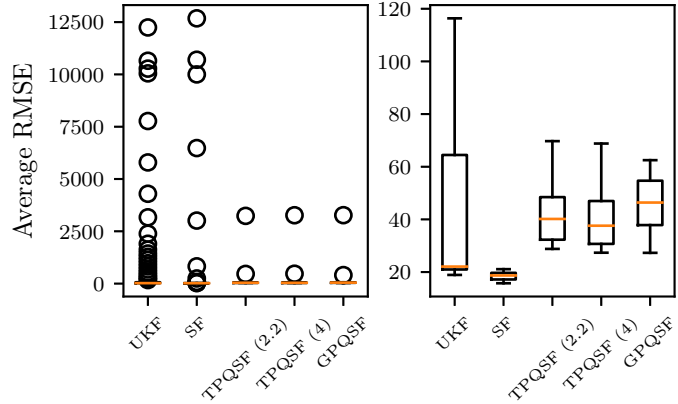}
	\caption{Overall filter RMSE shown with outliers (left) and a detail without outliers (right). The proposed TPQ-based filters have less extreme outliers, whereas the median RMSE favours the classical quadrature-based SF.}
	\label{fig:radar_rmse_boxplot}
\end{figure}

\begin{figure}[!htb]
	\centering
	\input{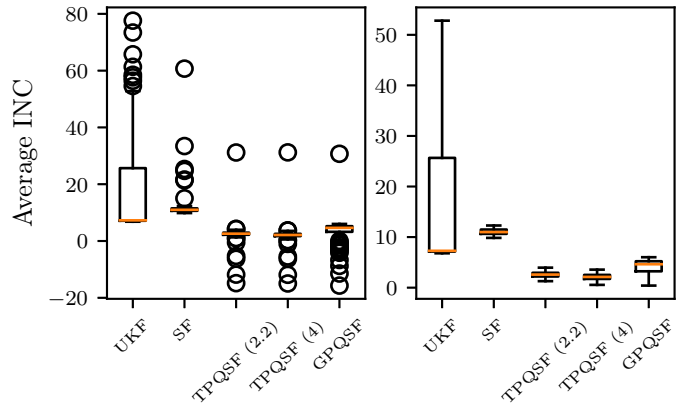}
	\caption{Overall filter INC shown with (left) and without (right) outliers. The proposed TPQ-based filters display improved INC with most outliers in the pessimistic direction, whereas the UKF and SF are excessively optimistic.}
	\label{fig:radar_inc_boxplot}
\end{figure}

The filter performance was evaluated by simulating 1,000 trajectories, each 100 time steps long, and computing the Monte Carlo averages of the performance scores.
\Cref{fig:radar_rmse_boxplot} shows box-plots of the time-averaged RMSE scores. 
The left pane shows that the UKF and SF have more extreme outliers than the proposed TPQSF, while in the right pane we see that the classical SF is better in terms of median RMSE.
It is worth noting that because TPQ-based filters have a tunable DoF parameter, they were able to achieve improved median RMSE over the GPQ-based filter.
From \Cref{fig:radar_inc_boxplot}, showing the time-averaged INC score, we can deduce that the BQ filters provide more balanced estimates on average, whereas the classical filters are excessively optimistic in their estimates.
This behaviour is in accordance with our expectations, because the BQ filters account for the additional functional uncertainty as described in \Cref{sec:general_tpq_transform}.
\Cref{tab:radar_rmse_inc} shows the mean of the overall average RMSE and INC along with the their standard deviations, which were estimated by bootstrapping with \num{10000} samples. 
Evidently, TPQSFs drastically improve the mean of the overall average RMSE and, as mentioned previously, provide much more balanced state estimates.
%

\begin{table}[!htb]
	\centering
	\begin{tabular}{lrrrr}
		\toprule
		{} 							&  RMSE 	&  STD 		&  INC 		&  STD \\
		\midrule
		UKF                			&   803.99 &        231.62 &   18.37 	&         1.80 \\
		SF                 			&   457.49 &        200.88 &   12.22 	&         0.58 \\
		TPQSF\( (\nlfDof=2.2) \) 	&    77.29 &         32.17 &    2.39 	&         0.38 \\
		TPQSF\( (\nlfDof=4) \)  	&    75.54 &         31.95 &    1.92 	&         0.39 \\
		GPQSF              			&    81.04 &         32.17 &    3.52 	&         0.45 \\
		\bottomrule
	\end{tabular}
	\caption{Overall RMSE and INC for the radar tracking example. The average RMSE favours the TPQ-based filters. 
		Our proposed filters also give more balanced state estimates on average, as shown by the inclination indicator (INC) being closer to zero.}
	\label{tab:radar_rmse_inc}
\end{table}

\section{Conclusion}\label{sec:conclusion}

In this article, we proposed a \student{} filter with a moment transform based on the \student{} process quadrature.
The MT is able to acknowledge the limited extent of knowledge of the integrated function when evaluated at finite number of evaluation points.
The proposed TPQ moment transform was applied in sigma-point \student{} filtering and its performance evaluated on two numerical examples.
Overall, the results indicate that the proposed TPQSF is superior at self-assessing its own performance and consequently provides more balanced estimates.

\section*{Acknowledgement}
Prüher and Straka were supported by the project LO1506 of the Czech Ministry of Education, Youth and Sports.
Tronarp and Särkkä were supported by Academy of Finland projects 266940, 273475 and 295080. 
Karvonen was supported by Aalto ELEC Doctoral School.

\bibliographystyle{plainnat}
\bibliography{bib/refs}
\end{document}